\documentclass[journal=jacsat,manuscript=article]{achemso}
\usepackage[version=3]{mhchem} 
\usepackage{hyperref}
\usepackage{xcolor}
\usepackage{array}
\usepackage{geometry}

\renewcommand\scriptsize{\fontsize{8.2}{8.2}\selectfont}
\author{Gareth A. Tribello}
\affiliation{Centre for Quantum Materials and Technology, School of Mathematics and Physics, Queen's University Belfast, UK}
\email{g.tribello@qub.ac.uk}
\author{Massimiliano Bonomi}
\affiliation{Institut Pasteur, Université Paris Cité, CNRS UMR 3528, Computational Structural Biology Unit, Paris, France}
\email{mbonomi@pasteur.fr}
\author{Giovanni Bussi}
\affiliation{Scuola Internazionale Superiore di Studi Avanzati (SISSA), Via Bonomea 265, 34136 Trieste, Italy}
\email{bussi@sissa.it}
\author{Carlo Camilloni}
\affiliation{Department of Biosciences, University of Milano, Milano, Italy}
\email{carlo.camilloni@unimi.it}
\author{Blake I. Armstrong}
\affiliation{School of Molecular and Life Sciences, Curtin University, GPO Box U1987, Perth, WA 6845, Australia}
\author{Andrea Arsiccio}
\affiliation{Coriolis Pharma, Fraunhoferstrasse 18b, 82152, Martinsried, Germany}
\author{Simone Aureli}
\affiliation{School of Pharmaceutical Sciences, University of Geneva, Rue Michel-Servet 1, CH-1206 Geneva, CH}
\alsoaffiliation{Institute of Pharmaceutical Sciences of Western Switzerland, University of Geneva, CH-1206, Geneva, CH}
\alsoaffiliation{Swiss Bioinformatics Institute, University of Geneva, CH-1206, Geneva, CH}
\author{Federico Ballabio}
\affiliation{Department of Biosciences, University of Milano, Milano, Italy}
\alsoaffiliation{National Research Council of Italy, Biophysics Institute (CNR-IBF), Via Celoria 26, Milan, 20133, Italy}
\author{Mattia Bernetti}
\affiliation{Department of Biomolecular Sciences, University of Urbino “Carlo Bo”, Piazza Rinascimento 6, 61029, Urbino, Italy}
\alsoaffiliation{Department of Pharmacy and Biotechnology, Alma Mater Studiorum - University of Bologna, Via Belmeloro 6, 40126 Bologna, Italy}
\author{Luigi Bonati}
\affiliation{Atomistic Simulations, Italian Institute of Technology, Genova, Italy}
\author{Samuel G. H. Brookes}
\affiliation{Yusuf Hamied Department of Chemistry, University of Cambridge, CB2 1EW Cambridge, UK}
\author{Z. Faidon Brotzakis}
\affiliation{Institute for Bioinnovation, Biomedical Sciences Research Center “Alexander Fleming”, 16672 Vari, Greece}
\alsoaffiliation{Yusuf Hamied Department of Chemistry, University of Cambridge, CB2 1EW Cambridge, UK}
\author{Riccardo Capelli}
\affiliation{Department of Biosciences, University of Milano, Milano, Italy}
\author{Michele Ceriotti}
\affiliation{Laboratory of Computational Science and Modeling, Institut des Matériaux, École Polytechnique Fédérale de Lausanne, 1015 Lausanne, CH}
\author{Kam-Tung Chan}
\affiliation{Department of Chemistry, University of California, Davis, Davis, CA 95616,
USA}
\author{Pilar Cossio}
\affiliation{Center for Computational Mathematics, Flatiron Institute, New York, New York 10010, USA}
\alsoaffiliation{Center for Computational Biology, Flatiron Institute, New York, New York 10010, USA.}
\author{Siva Dasetty}
\affiliation{Pritzker School of Molecular Engineering, University of Chicago, Chicago, Illinois 60637, USA}
\author{Davide Donadio}
\affiliation{Department of Chemistry, University of California, Davis, Davis, CA 95616,
USA}
\author{Bernd Ensing}
\affiliation{AI4Science Lab, Informatics Institute, University of Amsterdam, Amsterdam, Science Park 904, 1098 XH, The Netherlands}
\alsoaffiliation{Computational Chemistry Group, Van’t Hoff Institute for Molecular Sciences, University of Amsterdam, Amsterdam, Science Park 904, 1098 XH, The Netherlands}
\author{Andrew L. Ferguson}
\affiliation{Pritzker School of Molecular Engineering, University of Chicago, Chicago, Illinois 60637, USA}
\author{Guillaume Fraux}
\affiliation{Laboratory of Computational Science and Modeling, Institut des Matériaux, École Polytechnique Fédérale de Lausanne, 1015 Lausanne, CH}
\alsoaffiliation{Department of Chemistry, University of Chicago, Chicago, Illinois 60637, USA}
\author{Julian D. Gale}
\affiliation{School of Molecular and Life Sciences, Curtin University, GPO Box U1987, Perth, WA 6845, Australia}
\author{Francesco Luigi Gervasio}
\affiliation{School of Pharmaceutical Sciences, University of Geneva, Rue Michel-Servet 1, CH-1206 Geneva, CH}
\alsoaffiliation{Institute of Pharmaceutical Sciences of Western Switzerland, University of Geneva, CH-1206, Geneva, CH}
\alsoaffiliation{Swiss Bioinformatics Institute, University of Geneva, CH-1206, Geneva, CH}
\alsoaffiliation{Chemistry Department, University College London (UCL), WC1E 6BT, London, UK}
\author{Toni Giorgino}
\affiliation{National Research Council of Italy, Biophysics Institute (CNR-IBF), Via Celoria 26, Milan, 20133, Italy}
\author{Nicholas S. M. Herringer}
\affiliation{Department of Chemistry, University of Chicago, Chicago, Illinois 60637, USA}
\author{Glen M. Hocky}
\affiliation{Department of Chemistry, Simons Center for Computational Physical Chemistry, New York University, New York, NY, USA}
\author{Samuel E. Hoff}
\affiliation{Institut Pasteur, Université Paris Cité, CNRS UMR 3528, Computational Structural Biology Unit, Paris, France}
\author{Michele Invernizzi}
\affiliation{Peptone Ltd., The Connolly Works 41-43 Chalton Street London NW1 1JD, UK}
\author{Olivier Languin-Catto\"en}
\affiliation{Scuola Internazionale Superiore di Studi Avanzati (SISSA), Via Bonomea 265, 34136 Trieste, Italy}
\author{Vanessa Leone}
\affiliation{Department of Biophysics and Data Science Institute, Medical College of Wisconsin, Milwaukee, Wisconsin 53226-3548, USA}
\author{Vittorio Limongelli}
\affiliation{Euler Institute, Faculty of Biomedical Sciences, Università della Svizzera italiana (USI), via G. Buffi 13, CH-6900 Lugano, CH}
\author{Olga Lopez-Acevedo}
\affiliation{Biophysics of Tropical Diseases Max Planck Tandem Group, University of Antioquia UdeA, 050010 Medellin, Colombia}
\author{Fabrizio Marinelli}
\affiliation{Department of Biophysics and Data Science Institute, Medical College of Wisconsin, Milwaukee, Wisconsin 53226-3548, USA}
\author{Pedro Febrer Martinez}
\affiliation{School of Pharmaceutical Sciences, University of Geneva, Rue Michel-Servet 1, CH-1206 Geneva, CH}
\alsoaffiliation{Institute of Pharmaceutical Sciences of Western Switzerland, University of Geneva, CH-1206, Geneva, CH}
\alsoaffiliation{Swiss Bioinformatics Institute, University of Geneva, CH-1206, Geneva, CH}
\author{Matteo Masetti}
\affiliation{Department of Pharmacy and Biotechnology, Alma Mater Studiorum - University of Bologna, Via Belmeloro 6, 40126 Bologna, Italy}
\author{Shams Mehdi}
\affiliation{Department of Chemistry and Biochemistry, Institute for Physical Science and Technology, University of Maryland, College Park, MD, 20742, USA}
\affiliation{Department of Pharmacy and Biotechnology, Alma Mater Studiorum - University of Bologna, Via Belmeloro 6, 40126 Bologna, Italy}
\author{Angelos Michaelides}
\affiliation{Yusuf Hamied Department of Chemistry, University of Cambridge, CB2 1EW Cambridge, UK}
\author{Mhd Hussein Murtada}
\affiliation{Centre for Misfolding Diseases, Yusuf Hamied Department of Chemistry, University of Cambridge, Cambridge, UK}
\author{Michele Parrinello}
\affiliation{Atomistic Simulations, Italian Institute of Technology, Genova, Italy}
\author{Pablo M. Piaggi}
\affiliation{CIC nanoGUNE BRTA, Tolosa Hiribidea 76, 20018 Donostia-San Sebastián, Spain \& Ikerbasque, Basque Foundation for Science, 48013 Bilbao, Spain}
\author{Adriana Pietropaolo}
\affiliation{Dipartimento di Scienze della Salute, Università Magna Graecia di Catanzaro, Viale Europa, 88100 Catanzaro, Italy}
\author{Fabio Pietrucci}
\affiliation{Institut de Minéralogie, de Physique des Matériaux et de Cosmochimie, UMR 7590 CNRS, Sorbonne Université, Muséum National d'Histoire Naturelle, Paris 75005, France}
\author{Silvio Pipolo}
\affiliation{UCCS Unité de Catalyse et Chimie du Solide, Université de Lille, Université d’Artois UMR 8181, F-59000, Lille, France}
\author{Claire Pritchard}
\affiliation{PASTEUR, Département de chimie, École Normale Supérieure, PSL University, Sorbonne Université, CNRS, 24 rue Lhomond, 75005 Paris, France}
\author{Paolo Raiteri}
\affiliation{School of Molecular and Life Sciences, Curtin University, GPO Box U1987, Perth, WA 6845, Australia}
\author{Stefano Raniolo}
\affiliation{Euler Institute, Faculty of Biomedical Sciences, Università della Svizzera italiana (USI), via G. Buffi 13, CH-6900 Lugano, CH}
\author{Daniele Rapetti}
\affiliation{Scuola Internazionale Superiore di Studi Avanzati (SISSA), Via Bonomea 265, 34136 Trieste, Italy}
\author{Valerio Rizzi}
\affiliation{School of Pharmaceutical Sciences, University of Geneva, Rue Michel-Servet 1, CH-1206 Geneva, CH}
\alsoaffiliation{Institute of Pharmaceutical Sciences of Western Switzerland, University of Geneva, CH-1206, Geneva, CH}
\alsoaffiliation{Swiss Bioinformatics Institute, University of Geneva, CH-1206, Geneva, CH}
\author{Jakub Rydzewski}
\affiliation{Institute of Physics, Faculty of Physics, Astronomy and Informatics, Nicolaus Copernicus University, Grudziadzka 5, 87-100 Toruń, Poland}
\author{Matteo Salvalaglio}
\affiliation{Thomas Young Centre and Department of Chemical Engineering, University College London, London WC1E 7JE, UK}
\author{Christoph Schran}
\affiliation{Cavendish Laboratory, Department of Physics, University of Cambridge, Cambridge, CB3 0HE, UK}
\author{Aniruddha Seal}
\affiliation{Department of Chemistry, University of Chicago, Chicago, Illinois 60637, USA}
\author{Armin Shayesteh Zadeh}
\affiliation{Pritzker School of Molecular Engineering, University of Chicago, Chicago, Illinois 60637, USA}
\author{Tom\'as F. D. Silva}
\affiliation{Scuola Internazionale Superiore di Studi Avanzati (SISSA), Via Bonomea 265, 34136 Trieste, Italy}
\author{Vojt\v ech Spiwok}
\affiliation{University of Chemistry and Technology, Technická 5, CZ-166 28 Praha 6, Czech Republic}
\author{Guillaume Stirnemann}
\affiliation{PASTEUR, Département de chimie, École Normale Supérieure, PSL University, Sorbonne Université, CNRS, 24 rue Lhomond, 75005 Paris, France}
\author{Daniel Sucerquia}
\affiliation{Heidelberg Institute for Theoretical Studies, 69118 Heidelberg, Germany}
\author{Pratyush Tiwary}
\affiliation{Department of Chemistry and Biochemistry, Institute for Physical Science and Technology, University of Maryland, College Park, MD, 20742, USA}
\author{Omar Valsson}
\affiliation{Department of Chemistry, University of North Texas, Denton, TX, 76201, USA}
\author{Michele Vendruscolo}
\affiliation{Centre for Misfolding Diseases, Yusuf Hamied Department of Chemistry, University of Cambridge, Cambridge, UK}
\author{Gregory A. Voth}
\affiliation{Department of Chemistry, Chicago Center for Theoretical Chemistry, Institute for Biophysical Dynamics, and James Frank Institute, University of Chicago, Chicago, IL 60637, USA}
\author{Andrew D. White}
\affiliation{Department of Chemical Engineering, University of Rochester, Rochester, NY, USA}
\author{Jiangbo Wu}
\affiliation{Department of Chemistry, Chicago Center for Theoretical Chemistry, Institute for Biophysical Dynamics, and James Frank Institute, University of Chicago, Chicago, IL 60637, USA}

\title[PLUMED Tutorials]
  {PLUMED Tutorials: a collaborative, community-driven learning ecosystem}
\abbreviations{IR,NMR,UV}
\begin{document}

\newpage
\newgeometry{top=0.8in,bottom=0.8in,right=0.8in,left=0.8in}
\pagenumbering{arabic}
\setcounter{page}{2} 
\begin{abstract}
In computational physics, chemistry, and biology, the implementation of new techniques in a shared and open source software lowers barriers to entry and promotes rapid scientific progress.  However, effectively training new software users presents several challenges. Common methods like direct knowledge transfer and in-person workshops are limited in reach and comprehensiveness. Furthermore, while the COVID-19 pandemic highlighted the benefits of online training, traditional online tutorials can quickly become outdated and may not cover all the software's functionalities. To address these issues, here we introduce ``PLUMED Tutorials'', a collaborative model for developing, sharing, and updating online tutorials. This initiative utilizes repository management and continuous integration to ensure compatibility with software updates. Moreover, the tutorials are interconnected to form a structured learning path and are enriched with automatic annotations to provide broader context. This paper illustrates the development, features, and advantages of PLUMED Tutorials, aiming to foster an open community for creating and sharing educational resources.
\end{abstract}


\section{Introduction}

In fields like computational physics, chemistry, and biology, software is increasingly playing a central role in disseminating new methods and communicating scientific ideas. One such example is PLUMED (\href{https://www.plumed.org}{https://www.plumed.org}) \cite{plumed1,plumed2}, a popular open-source library that implements a variety of methods for enhanced-sampling, free-energy calculations, and analysis of molecular dynamics trajectories. Over the past 15 years, one of the central challenges that we, as members of the PLUMED community, have faced is how to effectively train new generations of scientists to use the methods implemented in our software. This challenge is common across different software in the field and becomes even more significant as the range of implemented functionalities increases.
 
Training new users usually involves direct knowledge transfer to students from supervisors or other members of a research group, often through inheriting old protocols and scripts. While effective for individual training, this approach has limited bandwidth, and requires a significant effort that ultimately only reaches one person. A complement to this model are in-person workshops and schools, where a larger audience of students receives training from experts in the field through a combination of theoretical lectures and practical exercises. However, the material delivered in these sessions is inevitably incomplete, as it is challenging for a few individuals to cover all relevant work in the field within the limited time available during in-person events. Furthermore, the recent COVID-19 pandemic and concerns about climate change have forced the global scientific community to reconsider the value of in-person meetings \cite{SCHWARZ2020101684}. Out of necessity, the community has realized that many traditional activities in scientific workshops such as lectures and exercises that students work through as individuals can be replicated online.  
 
In this context, online tutorials, which have always been a cornerstone for teaching researchers how to use methods implemented in software like PLUMED, have become even more indispensable. However, despite their undeniable educational value, the traditional format of online tutorials has several limitations. They are typically prepared by a few developers or experienced users and thus offer a limited overview of the methods implemented in the software. Moreover, as most online tutorials are delivered as static text, they can quickly become obsolete as the software evolves. 
This issue is similar to what we experienced with PLUMED tutorials that we published in journals and books in the past \cite{metad-book, Barducci2015, Löhr2019, Bussi2019, dimred1,dimred-2, clusters}.  Notable initiatives that aim to address these limitations are the “Tutorials and Training Articles” published in the Living Journal of Computational Molecular Science”\footnote{\url{https://livecomsjournal.org/index.php/livecoms/catalog/category/tutorials}}, the education resources of the The Molecular Sciences Software Institute (MolSSI)\footnote{\url{https://education.molssi.org/resources.html}}, the collection of documentation and best practice guides for BioExcel software\footnote{\url{http://docs.bioexcel.eu}}, and the tutorials developed and maintained by the worldwide Galaxy community\footnote{\url{https://training.galaxyproject.org}}.

Repository management systems, such as Git, coupled with continuous integration infrastructures, offer remedies to some of the issues described above as they facilitate collaborative development, versioning, and testing. These technologies have allowed us to design PLUMED as community software that researchers can both use and develop. Furthermore, these tools were instrumental when the PLUMED consortium was established in 2019\cite{plumedc}. This open community aims to transform how researchers communicate the protocols used in MD simulations through PLUMED-NEST (\href{https://www.plumed-nest.org}{https://www.plumed-nest.org}), an online repository for sharing PLUMED input files and thus facilitating the reproduction of results in published articles. However, PLUMED-NEST records are typically associated with published research and are not organized to provide a learning path. In fact, they are not intended to serve as tutorials.
 
Motivated by these considerations and building on our previous experience, here we present ``PLUMED Tutorials", a new model to develop online tutorials collaboratively. In this paper, we begin by examining the limitations of past PLUMED schools and explaining how our new initiative emerged from these previous experiences. We will then describe how PLUMED Tutorials has been designed to allow authors to easily create, share, and update their own tutorials. We will outline our process for regularly testing tutorial inputs to ensure they remain compatible with code changes. Additionally, we will discuss how PLUMED’s documentation is structured to automatically generate annotations that provide broader context for the example input files in each tutorial. We will also show how our tutorials can be interconnected to create a clear mastery path for new users. Finally, we will demonstrate the features that we have added for searching and indexing the tutorials. Importantly, our community is open, and we hope this paper will serve as a tutorial for creating and sharing new PLUMED tutorials.

\section{A brief history of PLUMED schools}

In-person schools, funded by CECAM and SISSA, were organized by the PLUMED developers in 2010\footnote{\url{https://www.cecam.org/workshop-details/free-energy-calculations-with-plumed-864}},
2012\footnote{\url{https://sites.google.com/site/plumedmeeting/home}}
2014\footnote{\url{https://www.cecam.org/workshop-details/enhancing-molecular-simulations-with-plumed-539}}, 2017\footnote{\url{https://sites.google.com/view/plumed-meeting-2017}}, 2019\footnote{\url{https://www.cecam.org/workshop-details/open-source-software-for-enhanced-sampling-simulations-118}}, and 2023\footnote{\url{https://www.cecam.org/workshop-details/enhanced-sampling-methods-with-plumed-1200}}. Overall, these meetings were attended by more than 350 participants. Early editions consisted of activities for students to work through during the meeting (more information at \href{https://www.plumed.org/funding}{https://www.plumed.org/funding}). This material was then uploaded to the PLUMED website (\href{https://www.plumed.org}{https://www.plumed.org}) to ensure that students who could not attend the event had access to it. However, this early model presented significant problems. Firstly, the material posted online was usually incomplete as it was designed for students who were attending in-person and could benefit from theoretical lectures that were not available online. Additionally, the lectures at each iteration of the school focused on different, though largely overlapping, topics. These differences usually required adjusting the tutorial activities to align with the changes. As a result, the PLUMED website ended up with several slightly different tutorials covering similar material and examples\footnote{\url{https://www.plumed.org/doc-v2.9/user-doc/html/tutorials.html}}.

Over the years, other members of the PLUMED community were involved in the meeting to ensure that the material presented during the school adequately reflected the state of the art in the field. However, students attending the meetings were often in the early stages of learning these techniques, therefore some time spent covering the basics was still necessary. We experimented with having either the school organizers or external experts give lectures on basic theory.  We found that having the organizers deliver these lectures was usually preferable as it allowed the lectures to better align with the tutorial content. We thus eventually adopted a hybrid format  where students spent the first few days attending lectures on basic theory and completing tutorials prepared by the meeting organizers. The final days of the school were then organized like a conventional scientific conference, with discussions and invited talks from members of the PLUMED community on the latest developments in the field.  While this model worked reasonably well, we still felt constrained by the short duration of the meeting. This limitation meant that students had to run simple and computationally inexpensive simulations, which left us doubtful that they were adequately prepared to understand the scientific talks or to develop real applications using the methodologies they had learned.
  
Another in-person PLUMED school was planned for 2021, but it was canceled due to the COVID-19 pandemic. This gave us the opportunity to reflect on the limitations of in-person schools, particularly the limited time available for students to engage in practical activities, and the advantages of running events online. Online activities can be completed asynchronously, allowing us to propose more complex exercises and computationally intensive simulations. Furthermore, an online format makes it far easier to record lectures and post a complete account of the school activities for students who could not attend the live meeting. We therefore organized a series of online events, called “PLUMED Masterclass” (\href{https://www.plumed.org/masterclass}{https://www.plumed.org/masterclass}). Each of these 7 sessions involved an hour-long lecture on Zoom that introduced the basic theory. Students were then assigned homework, which involved applying the concepts from the lecture to solve practical exercises. They had a week to complete this homework, during which a dedicated Slack channel was available for students to ask the lecturer questions about their assignment. At the end of the week, the lecturer then gave a second lecture that provided solutions to the homework assignment and addressed live questions from the online participants.

The material prepared for PLUMED Masterclass was possibly better than anything we have developed for previous, in-person meetings. This improvement was mostly due to having the time to refine old tutorials based on student feedback. We were, in fact, so satisfied with the content of the first masterclass series that, when we considered running a second series in 2022, we quickly ruled out covering the basics again.  We felt that students could learn these basics by working through the previous masterclasses asynchronously. For this second series, we thus instead invited members of the PLUMED community to present masterclasses focused  on the techniques they use and develop. The collection of online tutorials that was produced for these two masterclass series offers students a more complete introduction to the subject than any in-person PLUMED school ever could. All the lectures are available on YouTube (\href{https://www.youtube.com/@plumedorg1402}{https://www.youtube.com/@plumedorg1402}) and have already been viewed 49500 times. Furthermore, at the recent 2023 in-person PLUMED school, students were asked to complete some of the PLUMED Masterclass prior to the meeting. Students could then spend the time working together on short research projects and discussing what they had learned from their interactions with instructors and colleagues. We found this experience rewarding and were amazed by what the students were able to achieve within the short time available during the meeting.
 
For PLUMED Tutorials, we built on the experience gained over the past 15 years to develop a new infrastructure that will enable researchers to create and share PLUMED  tutorials. In the following sections, we will present our initiative in detail. 

\section{Overview of PLUMED Tutorials}

The PLUMED Tutorials website can be freely accessed at \href{https://www.plumed-tutorials.org}{https://www.plumed-tutorials.org}. On this website, readers can find a general description of our initiative as well as a list of available tutorials. This list is presented in a searchable table, which displays the name of the tutorial, the authors, and a short description. Currently, PLUMED Tutorials includes 47 contributions (Table \ref{table:1}). Among the topics, these tutorials cover instructions on how to install the software, familiarize  with the PLUMED syntax, perform enhanced-sampling simulations with popular methods, such as umbrella sampling\cite{TORRIE1977187}, metadynamics\cite{metad} and replica exchange\cite{SUGITA1999141}, analyze metadynamics simulations with Metadynminer\cite{RJ-2022-057}, design machine learning collective variables (CVs) with the mlcolvar module\cite{mlcolvar,opes-mlearn}, and perform SAXS-guided molecular dynamics simulations with the hySAS module\cite{hySAS}. Some of these tutorials have been created by adapting the material developed for PLUMED Masterclass. Furthermore, a few tutorials dedicated to PLUMED developers and contributors are also available, for example about parallelizing the calculation of CVs with OpenMP and MPI, or writing coordination-based CVs with CUDA. Since PLUMED is community-developed software, tutorials more oriented towards code contributors will be welcomed in the future. Finally, detailed instructions on how to contribute a tutorial are provided. In the following sections, we outline the steps needed to prepare and share PLUMED Tutorials and highlight the functionalities available to contributors.

\begin{table}[h!]
\scriptsize
\centering
\begin{tabular}{||m{3em} m{10cm} m{3cm} m{0.8cm} ||} 
 \hline
 ID & Topic & Author & Ref. \\ [0.5ex] 
 \hline\hline
 \href{https://www.plumed-tutorials.org/lessons/20/001/data/NAVIGATION.html}{20.001} & Installing PLUMED          & G. Tribello  & - \\ [0.5ex]
 \href{https://www.plumed-tutorials.org/lessons/21/001/data/NAVIGATION.html}{21.001} & PLUMED syntax and analysis        & M. Bonomi  & - \\ [0.5ex]
 \href{https://www.plumed-tutorials.org/lessons/21/002/data/NAVIGATION.html}{21.002} & Statistical errors in MD          & G. Tribello  & - \\ [0.5ex]
 \href{https://www.plumed-tutorials.org/lessons/21/003/data/NAVIGATION.html}{21.003} & Umbrella Sampling                 & G. Bussi  & \cite{TORRIE1977187} \\ [0.5ex]
 \href{https://www.plumed-tutorials.org/lessons/21/004/data/NAVIGATION.html}{21.004} & Metadynamics                      & M. Bonomi  & \cite{metad} \\ [0.5ex]
 \href{https://www.plumed-tutorials.org/lessons/21/005/data/NAVIGATION.html}{21.005} & Replica exchange methods          & G. Bussi  & \cite{SUGITA1999141} \\ [0.5ex]
 \href{https://www.plumed-tutorials.org/lessons/21/006/data/NAVIGATION.html}{21.006} & Dimensionality reduction          & G. Tribello  & \cite{pathcv} \\ [0.5ex]
 \href{https://www.plumed-tutorials.org/lessons/21/007/data/NAVIGATION.html}{21.007} & Optimizing PLUMED performances    & M. Bonomi  & - \\ [0.5ex]
  \href{https://www.plumed-tutorials.org/lessons/22/001/data/NAVIGATION.html}{22.001} & Funnel Metadynamics   & S. Raniolo, V. Limongelli  & \cite{funnel} \\ [0.5ex]
 \href{https://www.plumed-tutorials.org/lessons/22/002/data/NAVIGATION.html}{22.002} & Analysis of PLUMED output by Metadynminer   & V. Spiwok  & \cite{RJ-2022-057} \\ [0.5ex]
 \href{https://www.plumed-tutorials.org/lessons/22/003/data/NAVIGATION.html}{22.003} & Rethinking Metadynamics using the OPES method   & M. Invernizzi  & \cite{opes_new} \\ [0.5ex]
 \href{https://www.plumed-tutorials.org/lessons/22/005/data/NAVIGATION.html}{22.005} & Machine learning collective variables with PyTorch   & L. Bonati  & \cite{mlcolvar} \\ [0.5ex]
 \href{https://www.plumed-tutorials.org/lessons/22/006/data/NAVIGATION.html}{22.006} & EDS module and Coarse-Grained directed simulations & G. Hocky,  A. White & \cite{eds} \\ [0.5ex]
 \href{https://www.plumed-tutorials.org/lessons/22/007/data/NAVIGATION.html}{22.007} & Learning and enhancing fluctuations along information bottleneck for automated enhanced sampling   & P. Tiwary  & \cite{SPIB} \\ [0.5ex]
 \href{https://www.plumed-tutorials.org/lessons/22/008/data/NAVIGATION.html}{22.008} & Modelling concentration-driven processes with PLUMED   & M. Salvalaglio  & \cite{CμMD} \\ [0.5ex]
 \href{https://www.plumed-tutorials.org/lessons/22/009/data/NAVIGATION.html}{22.009} & Using path collective variables to find reaction mechanisms in complex free energy landscapes  & B. Ensing  & \cite{bernd} \\ [0.5ex]
 \href{https://www.plumed-tutorials.org/lessons/22/010/data/NAVIGATION.html}{22.010} & Hamiltonian replica exchange with PLUMED and GROMACS   & G. Bussi  & \cite{rem-gro} \\ [0.5ex]
  \href{https://www.plumed-tutorials.org/lessons/22/011/data/NAVIGATION.html}{22.011} & Variationally Enhanced Sampling   & O. Valsson  & \cite{omar-new-ves} \\ [0.5ex]
 \href{https://www.plumed-tutorials.org/lessons/22/012/data/NAVIGATION.html}{22.012} & Free energy calculations in crystalline solids   & P. Piaggi  & \cite{piaggi} \\ [0.5ex]
 \href{https://www.plumed-tutorials.org/lessons/22/013/data/NAVIGATION.html}{22.013} & SASA module - The solvent accessible surface area of proteins as a collective variable, and the application of PLUMED for implicit solvent simulations   & A. Arsiccio  & \cite{sasa} \\ [0.5ex]
 \href{https://www.plumed-tutorials.org/lessons/22/015/data/NAVIGATION.html}{22.015} &  Mechanical pulling + FISST module & G. Stirnemann, G. Hocky  & \cite{fisst} \\ [0.5ex]
 \href{https://www.plumed-tutorials.org/lessons/22/017/data/NAVIGATION.html}{22.017} &  A Bayesian approach to integrate cryo-EM data into MD simulations & S. Hoff, M. Bonomi  & \cite{emmivox} \\ [0.5ex]
 \href{https://www.plumed-tutorials.org/lessons/23/001/data/NAVIGATION.html}{23.001} & Developments in PLUMED   & G. Tribello  & - \\ [0.5ex]
 \href{https://www.plumed-tutorials.org/lessons/23/002/data/NAVIGATION.html}{23.002} & Introduction to the PLUMED parallel features for developers   & D. Rapetti  & - \\ [0.5ex]
 \href{https://www.plumed-tutorials.org/lessons/23/003/data/NAVIGATION.html}{23.003} & Profiling, GPUs and PLUMED   & K. Bhardwaj & - \\ [0.5ex]
 \href{https://www.plumed-tutorials.org/lessons/23/004/data/NAVIGATION.html}{23.004} & Rewriting coordination CVs in CUDA   & D. Rapetti  & - \\ [0.5ex]
 \href{https://www.plumed-tutorials.org/lessons/24/001/data/NAVIGATION.html}{24.001} & hybrid Small Angle Scattering — hands-on guide   & F. Ballabio  & \cite{hySAS} \\ [0.5ex]
 \href{https://www.plumed-tutorials.org/lessons/24/002/data/NAVIGATION.html}{24.002} & Trans-Cis isomerization in the ground and excited states using PLUMED & A. Pietropaolo  & \cite{adriana} \\ [0.5ex]
 \href{https://www.plumed-tutorials.org/lessons/24/003/data/NAVIGATION.html}{24.003} & Benchmarking PLUMED  & D. Rapetti  & - \\ [0.5ex]
 \href{https://www.plumed-tutorials.org/lessons/24/004/data/NAVIGATION.html}{24.004} & Volume-based Metadynamics  & R. Capelli  & \cite{volume} \\ [0.5ex]
 \href{https://www.plumed-tutorials.org/lessons/24/005/data/NAVIGATION.html}{24.005} & Path integral metadynamics  & M. Ceriotti  & \cite{ipi} \\ [0.5ex]
 \href{https://www.plumed-tutorials.org/lessons/24/006/data/NAVIGATION.html}{24.006}  & Standard binding free energies from cylindrical restraints  & B. I. Armstrong, P. Raiteri,  J. D. Gale & \cite{gale} \\ [0.5ex]
 \href{https://www.plumed-tutorials.org/lessons/24/007/data/NAVIGATION.html}{24.007}  & Transition-Tempered Metadynamics & J. Wu and G. A. Voth & \cite{ttmetad} \\ [0.5ex]
 \href{https://www.plumed-tutorials.org/lessons/24/008/data/NAVIGATION.html}{24.008}  & Using the maze module & J.  Rydzewski & \cite{maze} \\ [0.5ex]
 \href{https://www.plumed-tutorials.org/lessons/24/009/data/NAVIGATION.html}{24.009}  & Multiple Walkers Metadynamics Simulations with a Reactive Machine Learning Interatomic Potential & K.-T. Chan, D. Donadio & - \\ [0.5ex]
 \href{https://www.plumed-tutorials.org/lessons/24/010/data/NAVIGATION.html}{24.010}  & Modelling mechanobiological processes & C. Pritchard, G. Stirnemann, G. Hocky & \cite{stirne} \\ [0.5ex]
 \href{https://www.plumed-tutorials.org/lessons/24/011/data/NAVIGATION.html}{24.011}  & Parameterization of Path CVs for drug-target binding & M. Bernetti, M. Masetti & \cite{drug} \\ [0.5ex]
 \href{https://www.plumed-tutorials.org/lessons/24/012/data/NAVIGATION.html}{24.012}  & Exploring Free Energy Surfaces with MACE-PLUMED Metadynamics & S.G.H. Brookes, C. Schran, A. Michaelides	 & -\\ [0.5ex]
 \href{https://www.plumed-tutorials.org/lessons/24/013/data/NAVIGATION.html}{24.013}  & Permutationally Invariant Networks for Enhanced Sampling (PINES) & N.S.M. Herringer, A. Seal, A. Shayesteh Zadeh, S. Dasetty, A. L. Ferguson & \cite{pines} \\ [0.5ex]
 \href{https://www.plumed-tutorials.org/lessons/24/014/data/NAVIGATION.html}{24.014}  & Alpha-Fold Metainference for structural ensemble prediction of a partially disordered protein & F. Brotzakis, H. Murtada, M. Vendruscolo	 & \cite{AFMI} \\ [0.5ex]
 \href{https://www.plumed-tutorials.org/lessons/24/015/data/NAVIGATION.html}{24.015}  & How to use the PLUMED PyCV plugin & D. Rapetti, T. Giorgino	 & \cite{Giorgino2019} \\ [0.5ex]
 \href{https://www.plumed-tutorials.org/lessons/24/016/data/NAVIGATION.html}{24.016}  & Host-Guest binding free energies using an automated OneOPES protocol & P. Febrer Martinez, V. Rizzi, S. Aureli, F. L. Gervasio	 & \cite{oneOPES} \\ [0.5ex]
 \href{https://www.plumed-tutorials.org/lessons/24/017/data/NAVIGATION.html}{24.017}  & Enhanced sampling for magnesium-RNA binding dynamics & O. Languin-Catto{\"e}n	 & - \\ [0.5ex]
 \href{https://www.plumed-tutorials.org/lessons/24/018/data/NAVIGATION.html}{24.018}  & Permutation Invariant Vector and Water Crystallisation & S. Pipolo, F. Pietrucci & \cite{pipolo} \\ [0.5ex]
 \href{https://www.plumed-tutorials.org/lessons/24/019/data/NAVIGATION.html}{24.019}  & ASE-PLUMED interface & D. Sucerquia, P. Cossio, O. Lopez-Acevedo & \cite{ASE} \\ [0.5ex]
 \href{https://www.plumed-tutorials.org/lessons/24/020/data/NAVIGATION.html}{24.020}  & An introduction to CpH-Metadynamics simulations & T. F. D. Silva & \cite{silva2024characterizing} \\ [0.5ex]
 \href{https://www.plumed-tutorials.org/lessons/24/021/data/NAVIGATION.html}{24.021}  & Setting up and analyzing bias-exchange metadynamics simulations & V. Leone, F. Marinelli & \cite{bexch} \\ [0.5ex]
\hline
\end{tabular}
\caption{List of PLUMED Tutorials as of November 2024. For each tutorial, the ID hyperlinked to the tutorial page, topic, author(s), and main reference publication are reported.}
\label{table:1}
\end{table}

\section{How to contribute a PLUMED Tutorial}

PLUMED Tutorials are built using markdown files, python notebooks, images, and YouTube videos. Creating a tutorial simply involves bundling these resources into an online zip archive using a repository such as GitHub (\url{https://github.com}) or Zenodo (\url{https://zenodo.org}). The PLUMED Tutorials pages are then constructed using a GitHub Actions workflow, which downloads the contributed archives and constructs a GitHub pages site from the downloaded material. Contributors can thus edit the tutorials on the PLUMED Tutorials site by editing the contents of their online zip archive. The next time the PLUMED Tutorials page is constructed their edited archive is downloaded and their modified version of the tutorial will thus be rendered on the page.  This is very similar to the mechanism that we used to construct the PLUMED-NEST repository and is designed to make editing tutorials seamless and to empower contributors to maintain the tutorials they contribute.

\begin{figure}
\centering
\includegraphics[width=\textwidth]{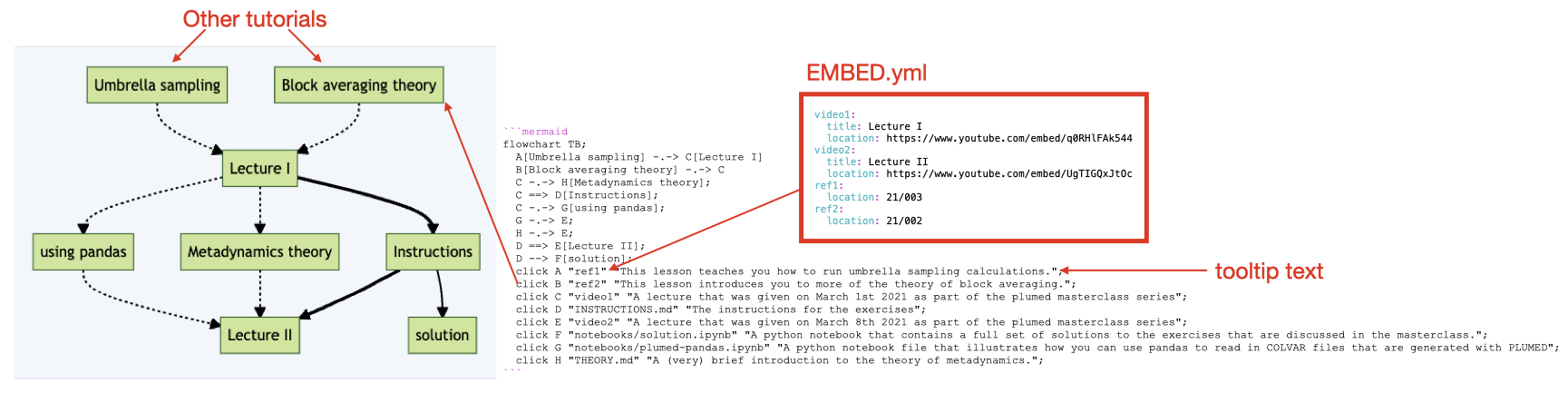}
\caption{An example of one of the graphs that show students how to work through a tutorial from our website and the markdown and yml text that is used to construct it.  As discussed in the text, when students hover over the green squares in the graph a tooltip appears that provides more detail on the resource that will open when they click on the square.  Instructions for creating the graph should be written in the NAVIGATION.md file in the Mermaid syntax that is shown on the right of figure.  If contributors want to include links to other tutorials on our site or external websites and videos they should include these links in an EMBED.yml file that is similar to the one shown here.}
\label{fig:fig1}
\end{figure}

On our site, a learner encountering a rendered tutorial sees a directed graph like that shown in the left panel of Fig. \ref{fig:fig1} which guides them through the various resources.  When the learner hovers over each node in the graph a tooltip appears that tells them a little about the corresponding resource.  Clicking on the node will then open the resource. For the tutorials we wrote we used a black solid line to indicate the path through the essential resources and dashed lines to indicate any supplementary resources that were provided. However, contributors also write a short paragraph of text on the landing page that explains the tutorial so they can choose to use line types in their graph in whatever way they feel is appropriate and describe their chosen approach.

The archive that contributors provide must contain a markdown file called NAVIGATION.md and a yaml file called EMBED.yml.  The landing page for the tutorial is constructed from the  NAVIGATION.md file, which should be written in GitHub markdown.  This file also contains the instructions for creating the directed graph that was described in the previous paragraph.  These graphs are created using Mermaid (\url{https://mermaid.js.org}), which is a JavaScript based diagramming and charting tool that takes Markdown-inspired text definitions and creates diagrams dynamically in the browser.  The right panel of Fig. \ref{fig:fig1} shows the Mermaid instructions for creating the diagram in the left panel of the figure.

When tutorial pages are constructed, a clickable resource is created every time a box in a Mermaid diagram that contains the work click is encountered.  For the example diagram in Fig. \ref{fig:fig1}, 8 resources are therefore created.  Resources can be created in one of the following six ways:

\begin{enumerate}
\item Markdown files can be provided in the zip archive. These files will then be rendered using GitHub pages. To construct links to these files you provide the path to the file in the archive when writing the instructions for constructing the Mermaid graph as shown for the item labeled Instructions in Fig. \ref{fig:fig1}.   
\item Python notebook files can be provided in the zip archive.  These files will then be rendered using GitHub pages.  To construct links to these files you provide the path to the file in the archive when writing the instructions for constructing the Mermaid graph as shown for the item labeled solution in Fig. \ref{fig:fig1}.   
\item PDF files can be provided in the  zip archive.  A page containing the embedded pdf will then be created by GitHub pages.  To construct links to these files you provide the path to the file in the archive when writing the instructions for constructing the Mermaid graph.   
Embedded videos can be included in tutorials.  If videos are included, a page with the embedded video is created when the tutorial is constructed.  Contributors provide the embed link for the video in the file EMBED.yml.  To construct a link to the page that contains the video you provide the keyword for the element in the yml dictionary that contains the video details in the instructions for constructing the Mermaid graph as shown for the items labeled Lecture I and II in Fig. \ref{fig:fig1}.  Pages with embedded videos are created when a title attribute is provided in the corresponding element of the yml dictionary.    
\item Links to resources that are external to the PLUMED Tutorials pages can be included.  These links are also provided in the EMBED.yml file.  When the contributor references the corresponding element of the yml dictionary in the instructions for constructing the Mermaid graph a link to the external website is created.  The contributor must, however, add the attribute ‘type: external’ to make it clear that the link in the corresponding element of the yml dictionary is to a page that is external to the PLUMED Tutorials website.
\item Links to other tutorials on the PLUMED Tutorials pages can be included.  Each tutorial has a code that appears on the main browse page.  To add a link to another tutorial the user adds an element to the EMBED.yml file that includes this code and the ‘type: internal’ attribute.  As shown for the items Umbrella Sampling and Block averaging theory in Fig. \ref{fig:fig1}, when the keyword for this dictionary element is included in the instructions for Mermaid, a link to the landing page for the tutorial that is referenced in the corresponding element of the yml dictionary is created.
\end{enumerate}

We believe that the six mechanisms for including tutorial content described above should cover the needs of our contributors.  However,  it is possible to extend the list above should the need arise in the future.  

Our aim is always to maximize flexibility for contributors so that the site reflects the diversity of views about these techniques that exist in the literature.  Contributors should be able to write tutorials even if they have not worked through all the other tutorials. Additionally, we do not think that having multiple tutorials on the same topic is necessarily to be avoided.  It would be nice if the site provided guidance to students on the order to work through tutorials.  However, we think this guidance is better provided by giving contributors a mechanism for linking other tutorials rather than imposing an order for users to work through the tutorials.  Ultimately, our site is designed in a way that more closely resembles a miniaturized version of the research literature than a textbook.  For the students who will use it, who are just embarking on a research career, this feels like an appropriate choice.

\section{Linking the tutorials with the documentation}

As discussed in the previous section, contributors can write their tutorials in GitHub markdown.  The ways in which hyperlinks, equations, code snippets, collapsible sections, Mermaid diagrams and images are included in such files is all thoroughly documented on the GitHub website.  The only other feature we have added to the GitHub markdown language for our PLUMED Tutorials website is a mechanism for including PLUMED input files.  Any PLUMED input code in a tutorial should be placed inside a block that starts with \verb_```plumed_ and that ends with \verb_```_.  In other words, PLUMED inputs are included in these markdown documents using a syntax that is similar to that used for including other snippets from other programming languages.

\begin{figure}
\centering
\includegraphics[width=0.6\textwidth]{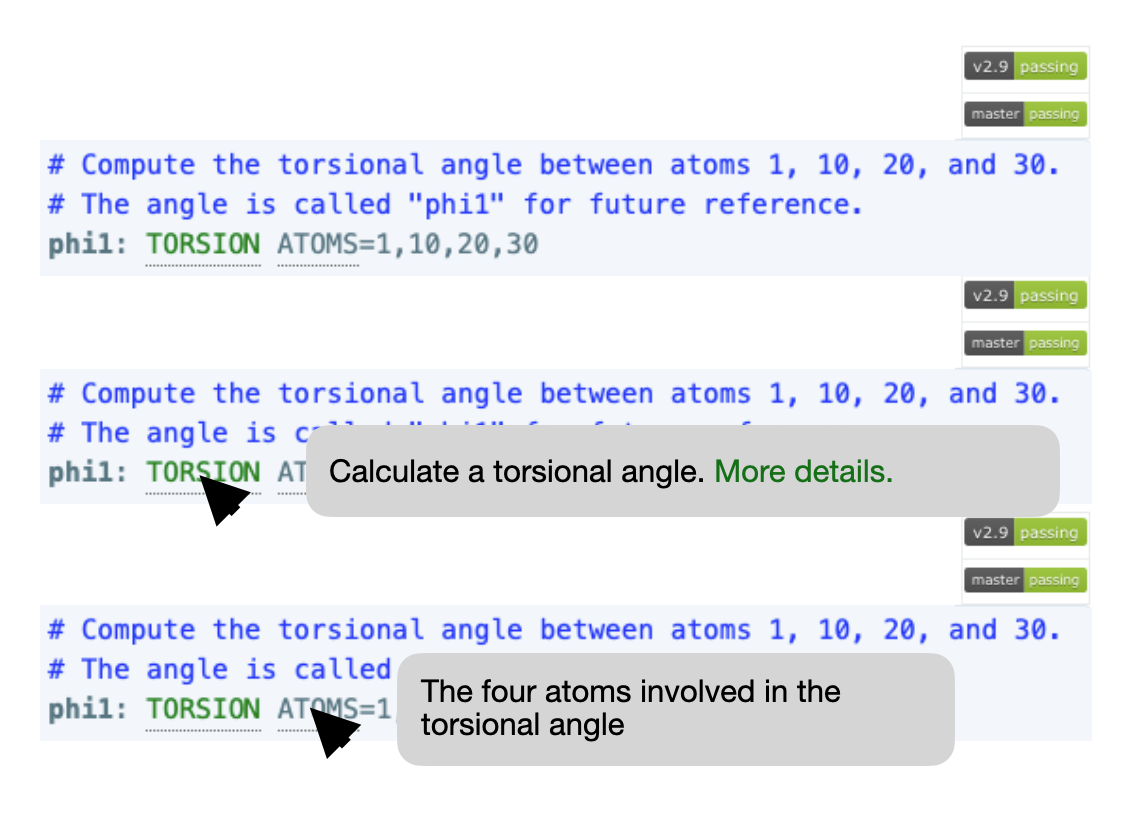}
\caption{An example showing how PLUMED input files are rendered when they appear on markdown pages in the tutorial.  The badges that appear on the upper right of the rendered input tell students whether the master and latest-stable version of PLUMED are still able to parse this input or not.  Furthermore, hovering over the keywords in the input file opens a tooltip that gives details on the keyword.  If the keyword is the name of the action then the tooltip that opens also contains a link to the relevant page in the manual.}
\label{fig:fig2}
\end{figure}

The code on the PLUMED Tutorials site that renders PLUMED input files provides more features than the code that GitHub uses to render other programming languages. For instance, when the renderer for our site encounters a block of PLUMED code it performs some basic tests to check that the instructions found are valid PLUMED input.  These checks are performed by determining whether the current release version and the master branch of the main development repository of PLUMED can parse the given input file.  As shown in Fig. \ref{fig:fig2}, if it is possible to parse the input then a green badge is shown beside the input.  If it is not possible to parse the input a red badge is shown.  These badges allow students to quickly identify tutorials that are obsolete and might need some work to be adapted to the current PLUMED version.  PLUMED developers can also use these badges to quickly identify the impacts of syntax changes for the community of researchers that are using PLUMED.

A useful technique that we have adopted when writing tutorials for PLUMED is to include gaps for students to complete in any example input files provided.  We have generally indicated these gaps using the keyword \verb^__FILL__^.  Gaps in input files are useful as they force students to look at the PLUMED documentation before running their calculations.  In other words, these gaps can be used to encourage students to think about the steps involved in a particular calculation rather than simply uncritically performing the tutorial instructions.  

The parser that generates the HTML input files is able to render PLUMED input files that contain gaps that are indicated using the \verb^__FILL__^ keyword. Furthermore, contributors can add a line containing the words \verb^#SOLUTIONFILE=solution.dat^ in the PLUMED input file that is provided in markdown to indicate a path where a complete version of the input with the gaps filled in can be found.  The colors of the badges that accompany the incomplete input file are decided by parsing the file that was provided to the \verb^#SOLUTIONFILE^ keyword using PLUMED.  Consequently, if PLUMED is able to parse the solution file that was provided, a green badge will appear by the input even though the input that appears in the markdown file is incomplete.  Finally, the \verb^#SOLUTIONFILE=solution.dat^ comment is not shown in the rendered input so there is nothing in the tutorial that indicates where the solution can be found.

PLUMED input files define a series of actions that must be performed.  Parameters that control how these actions are carried out are then specified using keyword value pairs.  As the lower panels in Fig. \ref{fig:fig2} show, hovering over each of the keywords in the rendered PLUMED input files on the PLUMED Tutorials site displays a tooltip that explains the purpose of each keyword.  If one hovers over the name of the action the tooltip that is shown contains a link to the page in the PLUMED manual that describes that action.

\begin{figure}
\centering
\includegraphics[width=\textwidth]{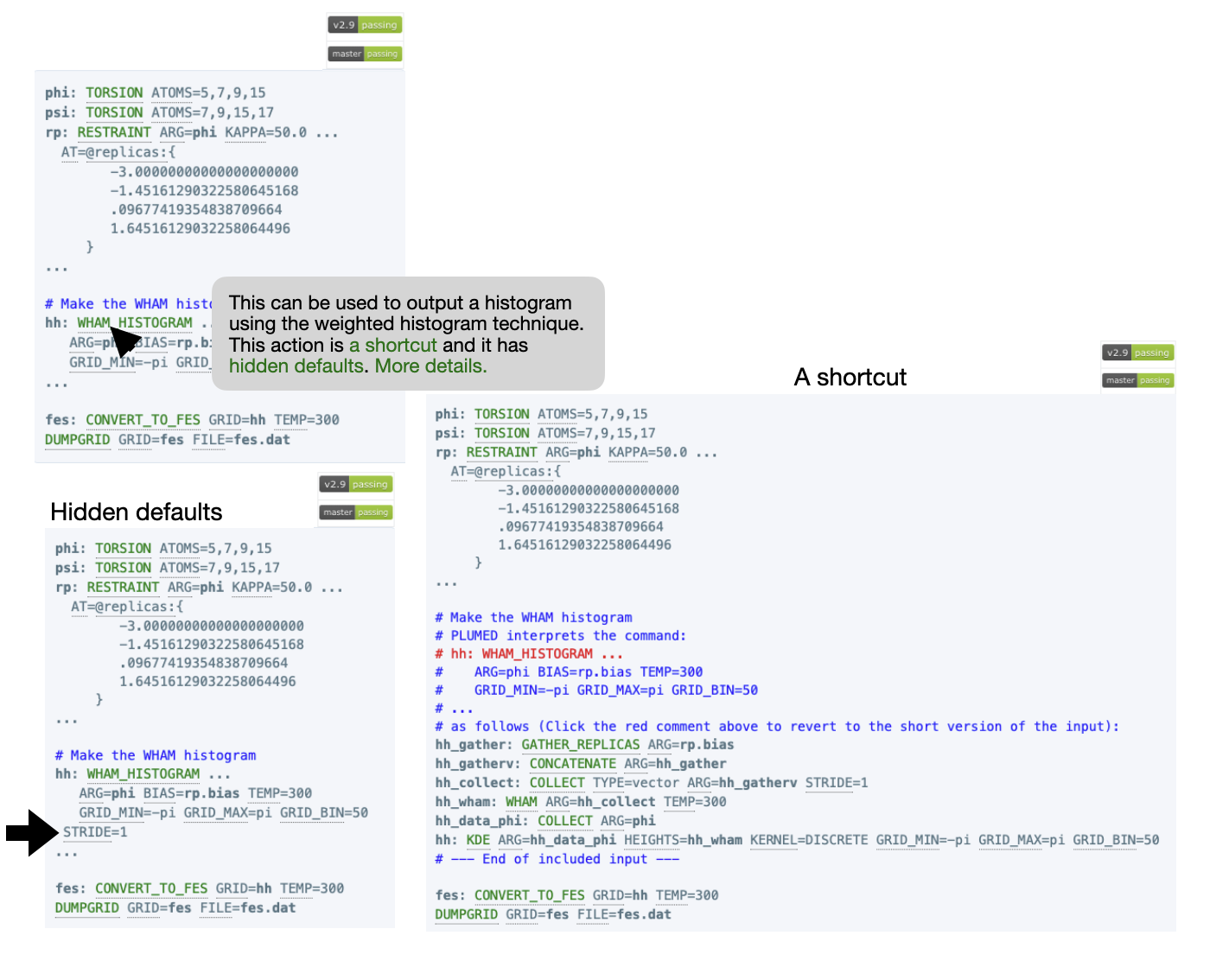}
\caption{An example of some of the advanced features of the rendered inputs that appear in the tutorial pages.  The tooltip for the WHAM\_HISTOGRAM command tells students that this command is a shortcut action and some of the parameter for this action are set to default values.  If the student clicks on the links in the tooltip the expanded input files shown in the bottom two panels are displayed.  Students can thus see the default values of the parameters and can also get some insight into how complex methods are implemented.}
\label{fig:fig3}
\end{figure}

The tooltips that appear when you hover over the names of some actions offer further links that can provide further detail on how actions operate.  For example, as we have already explained, the parameters of actions are set by providing keyword-value pairs.  In some cases, however, when a particular keyword is not present in the input line, the corresponding value is set to a hard-coded default.  As the bottom panel in Fig. \ref{fig:fig3} shows, when actions use these defaults, a show-defaults link appears in the action’s tooltip.  Clicking on this link reveals the keywords that would be used to set these parameters as well as the defaults that have been used in the calculation. 

A second type of link that can appear in the tooltip for an action is also shown in Fig. \ref{fig:fig3}.  In this case the action WHAM\_HISTOGRAM is a shortcut action.  These shortcut actions allow one to reuse and combine the functionality that is available in simpler PLUMED actions when performing more complex calculations.  In essence, a shortcut action is a wrapper that creates a complicated PLUMED input.  As the top panel in Fig. \ref{fig:fig3} shows, the tooltips for shortcut actions contain a shortcut hyperlink.  When the student clicks on this action the more complicated input that the shortcut creates is shown in the rendered PLUMED input as shown in the lower right panel of Fig. \ref{fig:fig3}.  Students can thus get a sense of how some of complex methods implemented in PLUMED are constructed by performing elementary mathematical operations such as matrix multiplication by expanding the shortcuts in the input files that appear in the tutorials on using such methods.

\begin{figure}
\centering
\includegraphics[width=\textwidth]{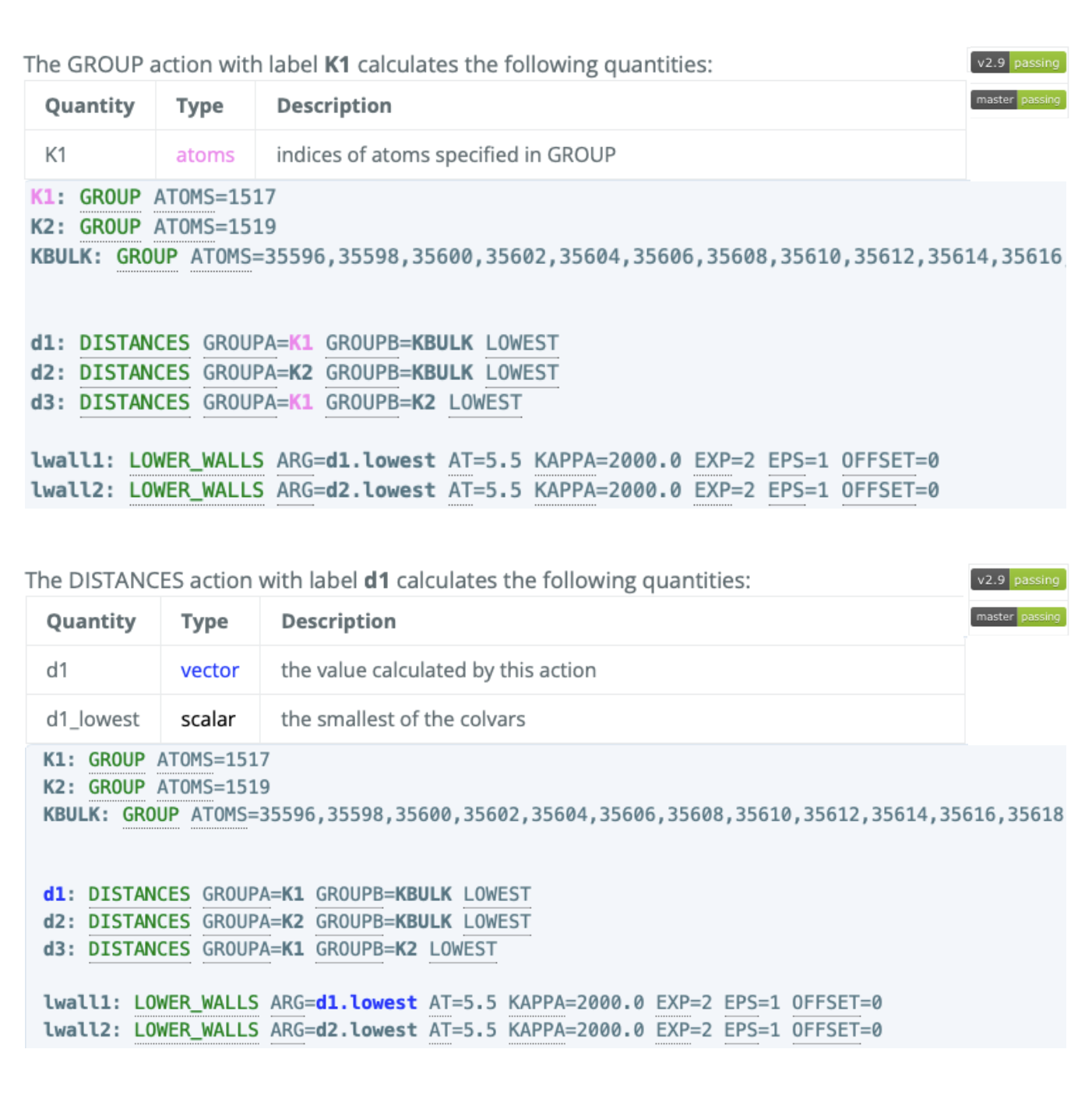}
\caption{An illustration showing how the values that are passed between actions are illustrated within PLUMED.  When students click on the label of an action a description of the values that are calculated by this action is displayed.  In addition, the places where the clicked-on value is used within the input is highlighted.}
\label{fig:fig4}
\end{figure}

PLUMED actions interact with each other by passing values.  These values can be atomic positions, scalars, vectors, matrices or functions evaluated on grids.  Fig. \ref{fig:fig4} shows how the role played by values is highlighted in the rendered inputs that are shown on tutorial pages.  As  Fig. \ref{fig:fig4} makes clear, when you click on the label of an action a description of the values it calculates is shown.  Additionally, the places where the values calculated by the action are used in the input are highlighted. The color that values are displayed in indicates the type of data that is being passed.  Violet is used to indicate the passing of atomic positions, black the passing of scalars, blue the passing of vectors, red the passing of matrices and green the passing of functions on grids. 

The tools for rendering the PLUMED input files described above have been incorporated into the code for building the PLUMED-NEST  pages as well as the tutorial pages.  We have thus rendered over 1600 example inputs using these tools.  We believe that these examples provide a much richer body of information on how to use PLUMED than a manual ever could.  We believe that students learn by taking example inputs that have been used in other projects and modifying them to a new purpose.  We hope that by annotating the inputs using the tools described above we have made the process of understanding what a particular input file does and how it might be modified to a new purpose more straightforward.

\section{Indexing the tutorials}

When a student opens the browse page of our website they are shown a table that contains the name of the tutorial, its author and a short description. The tutorials appear in reverse chronological order so the most recently submitted tutorials are shown first.  Students can then reorder the tutorials alphabetically by name or author name.

\begin{figure}
\centering
\includegraphics[width=\textwidth]{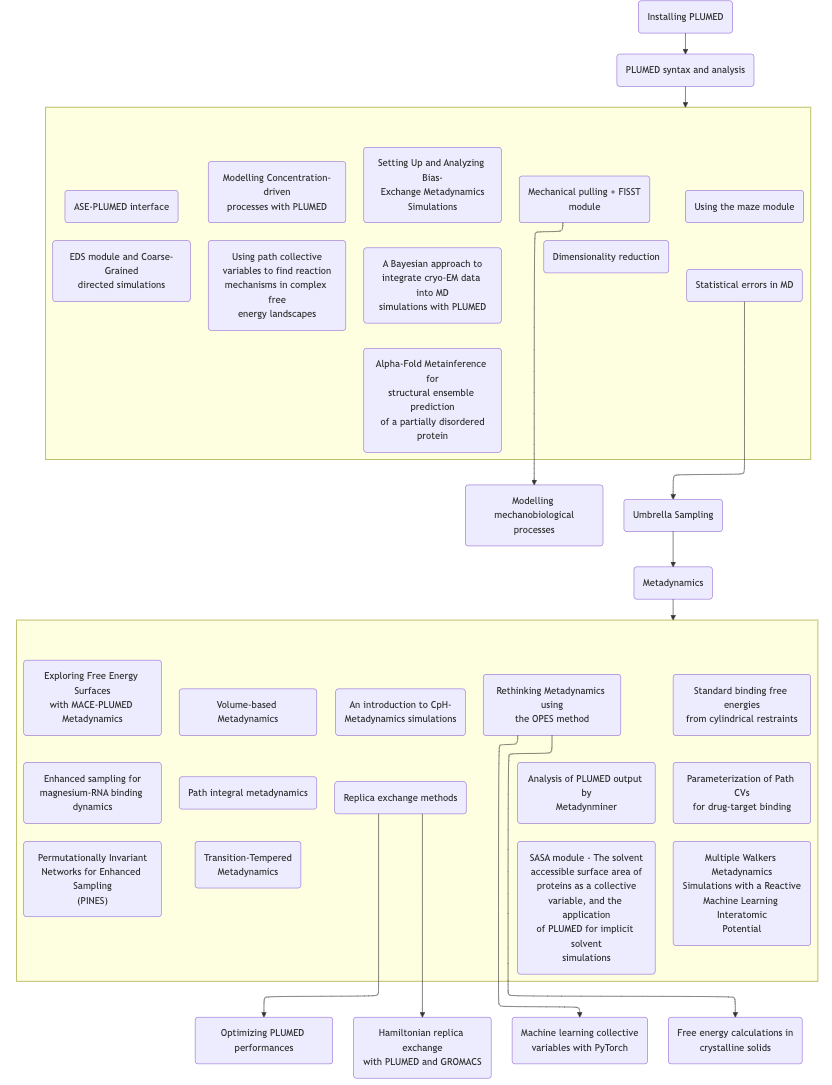}
\caption{Selected tutorials that are available on the PLUMED Tutorials website and the order that contributors have suggested students should work through them.}
\label{fig:5}
\end{figure}

Fig. \ref{fig:5} shows another representation of the tutorials that students can use to browse through them.  The graph in this figure gives students advice on the order in which students should work through the tutorials.  To construct it we use the links that contributors have included to other tutorials.  We create a directed graph from these dependencies, remove any circular dependencies and then perform a transitive reduction on the resulting tree \cite{transitive-reduction}. In this way the advice we provide on the order to work through the tutorials is democratically determined by contributors to the site.  Furthermore, providing this representation encourages contributors to think about how the material in their tutorial fits with the other available content as, if a contributor does not include links within their tutorial and if their tutorial is not linked by any other tutorial, it does not appear in the graph in Fig. \ref{fig:5}. 

The tutorials can be searched by typing a search term in the search bar on the browse page.  Alternatively, if authors wish to create link a particular subset of tutorials, for example if they wish add a link to the list of tutorials they have authored on their personal website, they can write a link to the browse page as follows:

\url{https://www.plumed-tutorials.org/browse.html?search=search_term}

A person clicking on this link is directed to a version of the browse page of the tutorials site that has the search bar populated with whatever the link's author used in place of \verb$search_term$. Multiple keys can be added using the html string ``\%20'' as a separator. So, for example, the link: \\ \url{https://www.plumed-tutorials.org/browse.html?search=tribello%20dimensionality} \\
opens a browser page that lists all the tutorials by Gareth Tribello on dimensionality reduction.  We believe providing functionality to search the school using a link is useful as it ensures that links to lists of relevant tutorials 
automatically update whenever new content is submitted to the site.          

It is also possible to search and find tutorials where a particular action or module is used.  A list of the actions used in a particular tutorial and the modules they are part of is built automatically when the input files that appear in the markdown file instructions are rendered.  This automatically-constructed list is then used when searching tutorials for actions. Consequently, when a contributor employs a METAD action in an example input it is automatically added to the list of the tutorials that appears when students search for METAD.  Links that open a list  of tutorials (and items in PLUMED-NEST) that use each of the actions/modules have also been made available within PLUMED are shown on the manual pages for those actions/modules. Students and contributors can thus understand how frequently and how each feature within PLUMED is used by reading the manual.   

\section{Conclusions}

PLUMED’s visibility online is the result of the efforts of a community of users and developers who have used the code and contributed to it. It is important to highlight the collaborative nature of this work as this is the main reason for the code’s success. Collective projects like these can only continue to succeed when every person involved feels that they are getting the credit they deserve for their contributions. 

Our new initiative, PLUMED Tutorials, results from years of experience as software users, developers, trainees, and trainers. We believe this resource addresses many limitations we encountered in previous PLUMED training activities and traditional tutorials. In particular, as PLUMED Tutorials are continuously built through a collaborative community effort, they offer a more comprehensive overview of the software's functionalities. Additionally, the automated testing of the tutorial inputs simplifies maintaining both the code and the tutorials. Similarly, we have shown how our site automates linking each tutorial with the wider body of code documentation through the automated generation of tooltips and hyperlinks.

Beyond training, PLUMED Tutorials can complement, rather than replace, journal articles by disseminating knowledge in ways that traditional publications cannot. These tutorials enhance efforts toward data sharing based on F.A.I.R. principles\cite{amaro2024needimplementfairprinciples} but focuses on reusing methods rather than  existing data sets.
The text in these tutorials may even prove useful when training large language models to run MD simulations.  Essentially, when an author writes a tutorial alongside a scientific article, it may encourage readers to engage more deeply with the paper, increasing the research's impact. For example, when we used the tutorial material to train students in our own research groups, we found that students were more inclined to adopt methods developed by other groups. Thus, PLUMED Tutorials will contribute to the cross-fertilization of ideas between research groups and encouraging a deeper engagement with the simulation techniques developed by our colleagues.

Finally, we foresee that PLUMED Tutorials will play a central role in future in-person PLUMED schools and activities. Students can learn basic concepts and practice using this resource before attending the meeting. This approach would free up time for other activities during the meeting, such as collaborative short projects and discussions with colleagues and instructors. Given the reality of climate change, conference and school organizers should provide a compelling case for why the event cannot be run online and explain the added value of bringing students and academics together in the same physical space. However, for this transition in organizing scientific meetings to occur, new shared online spaces must be developed to enable more knowledge exchange outside of in-person meetings. With PLUMED Tutorials, we aimed to develop such infrastructure, and we hope that developers of other software in the field will find this useful.

\begin{acknowledgement}
The authors thank all the participants and instructors of previous PLUMED in-person meetings for providing feedback on the format and content, as well as CECAM, SISSA, CSCS, and Schrödinger for providing financial support to these events.  M.B. acknowledges funding from the European Research Council (ERC) under the European Union’s Horizon 2020 research and innovation programme (“bAIes” ERC grant agreement no. 101086685). M.B. and S.E.H. acknowledge the support of the French Agence Nationale de la Recherche (ANR), under grant ANR-20-CE45-0002 (project EMMI). S.E.H. is founded by a Roux-Cantarini fellowship from the Institut Pasteur (Paris, France). S.G.H.B. was supported by the Syntech CDT and funded by EPSRC (Grant No. EP/S024220/1). P.M.P. acknowledges funding from the Marie Sklodowska Curie Cofund Programme of European Commission project H2020-MSCA-COFUND-2020101034228-WOLFRAM2.
O.L.-C. acknowledges the
Next Generation EU project PRIN 2022 (2022Z4FZE9).
T.S. acknowledges the
European Molecular Biology Organization (EMBO) through grant
ALTF-399/2022.
G.B. and D.R. acknowledge the Italian National Centre for HPC, Big Data, and Quantum Computing (grant No. CN00000013), founded within the Next Generation EU initiative.
V.L. and S.R. acknowledge funding from the European Research Council (ERC) under the European Union’s Horizon 2020 research and innovation programme (‘‘CoMMBi’’ ERC grant agreement no. 101001784) and the Swiss National Supercomputing Centre (CSCS) under project ID u8 and s1293.  J.D.G thanks the Australian Research Council for funding under grant FL180100087, as well as the Pawsey Supercomputing Centre and National Computational Infrastructure for resources.  C.S. acknowledges partial financial support from the Royal Society via grant number RGS\textbackslash R2\textbackslash 242614. GMH was supported by NIH grant R35GM138312.  K-T.C. and D.D. acknowledge support by the National Science Foundation under Grant No. 2305164. O.V. acknowledges funding from US Department of Energy, Office of Science, Basic Energy Sciences, CPIMS Program, under Award DE-SC0024283. M. B. and M. M. gratefully acknowledge Davide Branduardi for sharing preliminary scripts, insightful discussions and technical support.  The Flatiron Institute is a division of the Simons Foundation.
\end{acknowledgement}

\section{Conflict of Interest Disclosure}

A.L.F. is a co-founder and consultant of Evozyne, Inc. and a co-author of US Patent Applications 16/887,710 and 17/642,582, US Provisional Patent Applications 62/853,919, 62/900,420, 63/314,898, 63/479,378, 63/521,617, and 63/669,836, and International Patent Applications PCT/ US2020/035206, PCT/US2020/050466, and PCT/US24/10805.




\bibliography{achemso-demo}

\end{document}